\documentclass[10pt,journal,compsoc]{IEEEtran}
\usepackage{amsmath,amsfonts}
\usepackage{algorithmic}
\usepackage{algorithm}
\usepackage{array}
\usepackage[caption=false,font=normalsize,labelfont=sf,textfont=sf]{subfig}
\usepackage{textcomp}
\usepackage{stfloats}
\usepackage{url}
\usepackage{verbatim}
\usepackage{graphicx}

\usepackage{color}
\usepackage{multicol}
\usepackage{array}
\usepackage{stfloats}
\usepackage{amsthm} 
\usepackage{multirow}
\usepackage{amssymb}
\usepackage{float}
\usepackage{nccmath}
\usepackage{graphicx}
\usepackage{setspace}
\usepackage[algo2e, linesnumbered, ruled, vlined]{algorithm2e}
\usepackage{mathtools}
\usepackage{booktabs}
\newcommand{\thetab}{\ensuremath{\bm{\theta}}}
\usepackage{mwe}
\usepackage{cite}
\usepackage{amsmath,amssymb,amsfonts}
\usepackage{graphicx}
\usepackage{textcomp}
\usepackage{xcolor}
\usepackage{physics}
\usepackage{bm}
\usepackage[utf8]{inputenc} % use UTF8 encoding
\usepackage{kotex} % use KoTeX package for Korean 
\usepackage[linesnumbered,ruled]{algorithm2e}
\usepackage{cite}
\usepackage{amsmath,amssymb,amsfonts}
\usepackage{algorithmic}
\usepackage{graphicx}
\usepackage{textcomp}
\usepackage{xcolor}
\usepackage{mathtools}
\DeclarePairedDelimiter\ceil{\lceil}{\rceil}

\newcommand{\lcal}{\ensuremath{\mathcal{L}}}

\usepackage[normalem]{ulem}

\ifCLASSINFOpdf
\else
\fi
\begin{document}
% \title{Bare Demo of IEEEtran.cls for\\ IEEE Computer Society Journals}

\title{Scalable Quantum Convolutional Neural Networks}
% by Extracting Diverse Features

\author{
    Hankyul Baek,
    Won Joon Yun, 
    and
    Joongheon Kim,~\IEEEmembership{Senior Member, IEEE}
    \IEEEcompsocitemizethanks{
    \IEEEcompsocthanksitem This research was funded by the National Research Foundation of Korea (2022R1A2C2004869). 
    \IEEEcompsocthanksitem\textit{Hankyul Baek and Won Joon Yun contributed equally to this work (first authors). Joongheon Kim is the corresponding author of this paper.}
    \IEEEcompsocthanksitem {Hankyul Baek, Won Joon Yun, and Joongheon Kim are with the School of Electrical Engineering, Korea University, Seoul 02841, the Republic of Korea (e-mails: \{67back,ywjoon95,joongheon\}@korea.ac.kr).}}%
}

%\markboth{Journal of \LaTeX\ Class Files,~Vol.~XX, No.~X, Month~Year}%
%{Shell \MakeLowercase{\textit{et al.}}: Bare Demo of IEEEtran.cls for Computer Society Journals}
\IEEEtitleabstractindextext{%
\begin{abstract}
With the beginning of the noisy intermediate-scale quantum (NISQ) era, quantum neural network (QNN) has recently emerged as a solution for the problems that classical neural networks cannot solve. Moreover, QCNN is attracting attention as the next generation of QNN because it can process high-dimensional vector input. However, due to the nature of quantum computing, it is difficult for the classical QCNN to extract a sufficient number of features. Motivated by this, we propose a new version of QCNN, named scalable quantum convolutional neural network (sQCNN). In addition, using the fidelity of QC, we propose an sQCNN training algorithm named reverse fidelity training (RF-Train) that maximizes the performance of sQCNN.
\end{abstract}

}

\maketitle

\IEEEdisplaynontitleabstractindextext
\IEEEpeerreviewmaketitle

\IEEEraisesectionheading{\section{Introduction}\label{sec:introduction}}
\IEEEPARstart{Q}{uantum} computing is anticipated to outperform classical algorithms in processing speed and impact various industry sectors that need complex computation~\cite{Mobicom22Kim, TETC1,spl01, spl02, TETC2}. Based on the quantum entanglement theory~\cite{aimlab2022icdcs}, each constituent of a quantum system is an inseparable whole, and the state in a quantum system is expressible as a superposition of states of each local constituent. In contrast to classical computation, where the computation unit (\textit{i.e.}, bit) holds either 0 or 1, a computing unit in the quantum system (\textit{i.e.}, qubit) can represent a superposition of the states, $\alpha \ket{0} + \beta \ket{1}$, where $\alpha$ and $\beta$ are complex numbers~\cite{aimlab2022arvix,kong2020review}. This superposition of the states enables the quantum system to process a lot of vectorized data with only a small amount of qubits~\cite{TETC0, huang2021power}. The quantum system express and computes the data on an exponential scale. Therefore, even in the current decade of noisy intermediate-scale quantum computation (NISQ), quantum machine learning (QML) has acquired linear or sublinear complexity as opposed to the polynomial complexity of conventional ML. Since conventional ML depends heavily on massive data, which is extremely hard to analyze and process, QML has drawn attention as a practical solution to these challenges. Various research has been conducted to utilize the nature of quantum computing on QML fully. For instance, a classification task, one of the representative machine learning problems, can be solved by a QML-based classifier~\cite{blank2020quantum}. In addition, QML can be used with not only itself but classical neural networks~\cite{schuld2019quantum}. Previous research showed that quantum computing performs complex computations in Hilbert space more efficiently than classical computing~\cite{schuld2019quantum}.
However, there is still a challenging problem that QML faces,~\textit{i.e.}, barren plateaus. The barren plateaus are a notorious problem in QML that occurs when the number of qubits increases. The barren plateaus vanish the gradients of the QML, making it impossible to guarantee trainability~\cite{mcclean2018barren}. As a solution, research in~\cite{pesah2021absence} proved that utilizing a quantum convolutional neural network (QCNN) with proper initialization can reduce the barren plateaus. Recent research in~\cite{DBLP:journals/qmi/HendersonSPC20} designed the QCNN with filters to extract the features of input data like classical convolutional neural networks (CNN). Inspired by it, this paper proposes a new version of QCNN, \textit{scalable quantum convolutional neural network} (\textit{sQCNN}), and a new training algorithm,~\textit{reverse fidelity-train} (\textit{RF-Train}), which utilizes the concept of fidelity,~\textit{i.e.}, the nature of quantum computing. Therefore, sQCNN can fully use the intrinsic features of the input data. 

\begin{figure}[t!]
\centering
\includegraphics[width=\columnwidth]{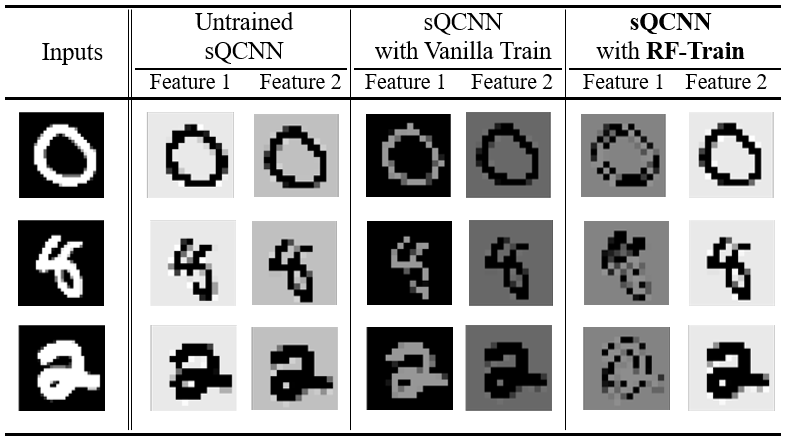}
\caption{Extracted feature maps according to various train strategies with sQCNN (untrained, vanilla trained, RF-trained).}\label{fig:Featuremap}
\end{figure}

\subsection{Contributions} The contributions of this research are summarized as follows.
\begin{itemize}
    \item First of all, we propose a scalable QCNN architecture with quantum computing, \textit{i.e.}, sQCNN, to achieve the scalability of filters while avoiding barren plateaus by maintaining QCNN architecture.
    
    \item In addition, we propose an sQCNN training algorithm (named \textit{RF-Train}) to extract the intrinsic features with finite filters.
    
    \item Lastly, we conduct data-intensive experiments to corroborate the superiority of sQCNN with RF-Train in MNIST and FMNIST datasets, widely used in the literature.  
\end{itemize}

Fig.~\ref{fig:Featuremap} shows each feature map of untrained QCNN, Vanilla-Trained QCNN, and RF-trained sQCNN, respectively. We describe each feature's classification performance and Euclidean distance from these models in Sec.~\ref{sec:experiment}.

\subsection{Organization} 
The rest of this paper is organized as follows.
Sec. \ref{sec:2} provides the descriptions on CNN and QCNN. 
After that, Sec. \ref{sec:3} describes the proposed sQCNN and its training algorithm; and Sec. \ref{sec:experiment} presents the performance evaluation results.
Finally, Sec. \ref{sec:5} concludes this paper.

\section{Preliminaries of sQCNN}\label{sec:2}

\subsection{Classical Convolutional Neural Network}
A classical CNN architecture is mainly composed by 3 layer components, \textit{i.e.}, convolution layers, pooling layers, and fully connected layers. 

\subsubsection{Convolution Layers} In convolution layers, input data are convolved by a set of filters. Each filter is designed to extract the intrinsic feature of the input data. The number of filters can be adjusted under the consideration of elapsed calculation time in each layer in this classical CNN. The output is called the feature map of this convolution computation.

\subsubsection{Pooling Layers}
In pooling layers, dimensionality reduction is conducted on the convolved data. This procedure is essential as it reduces the computation time on the next convolution layer. Moreover, it allows CNN to learn representations invariant to small translations.

\subsubsection{Fully Connected Layers}
After the computational procedure of convolution layers and pooling layers, conducting fully connected layers on input data enables the model to get the probability of initial input belonging to the corresponding class. 

\begin{figure*}[t!]
\centering
\includegraphics[width=0.85\linewidth]{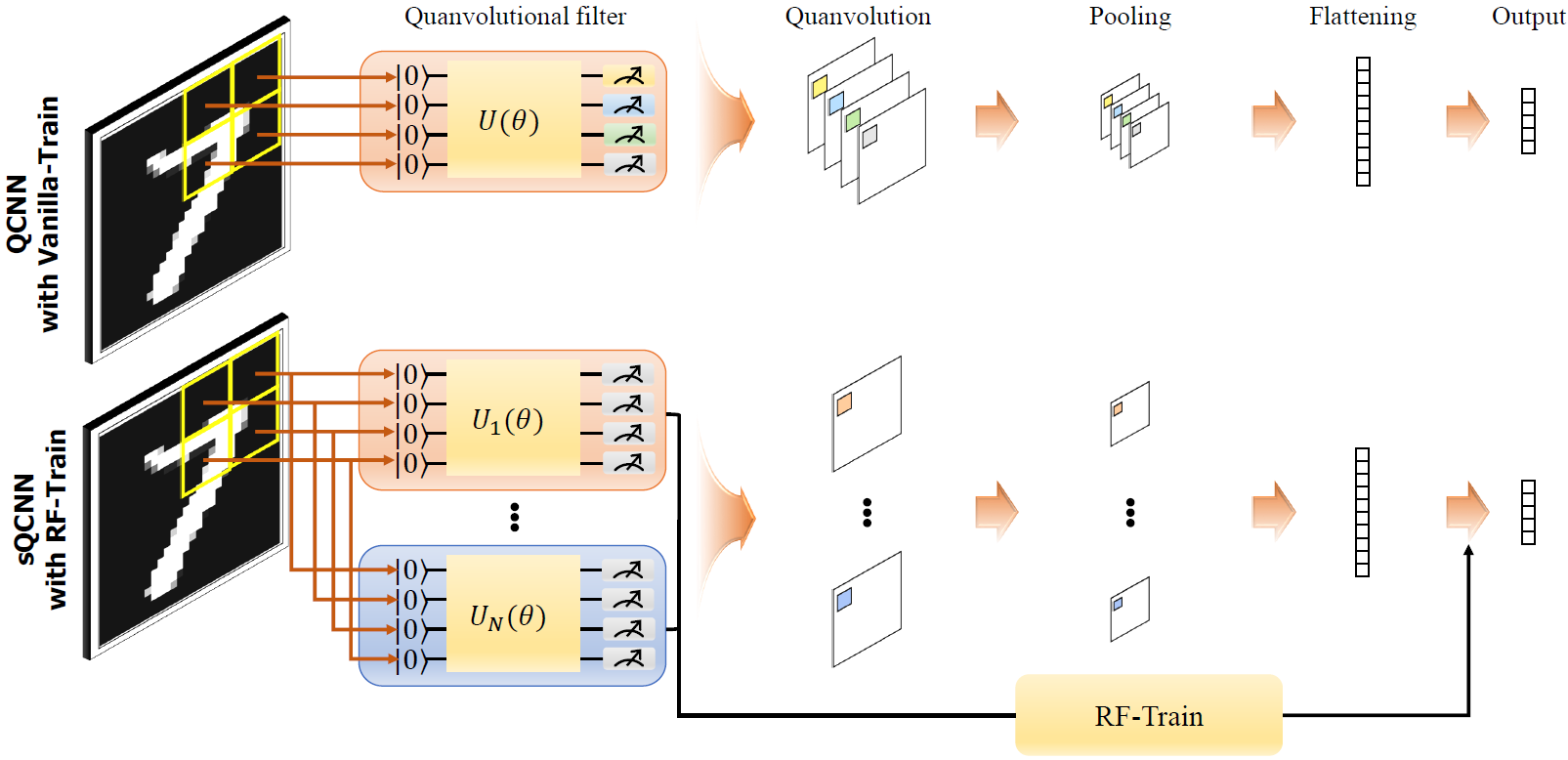}
\caption{Comparison between QCNN with Vanilla-Train and sQCNN with RF-Train.}\label{fig:Model ARchi}
\end{figure*}

\subsection{Quantum CNN}\label{sec:quanvolution}
QCNN is a new design of CNNs for multi-dimensional vectors using quantum circuits as convolutional filters~\cite{DBLP:journals/qmi/HendersonSPC20}. To consider spatial information with quantum computing, QCNN adopts a quantum version of convolution (\textit{i.e.}, quanvolution)~\cite{DBLP:journals/qmi/HendersonSPC20}. The quanvolutional filter consists of three components,~\textit{i.e.}, encoder, parameterized quantum circuit (PQC), and measurement. The architecture of quanvolutional filter is illustrated in Fig.~\ref{fig:Model ARchi}. The encoder enables the quanvolutional filter train with the classical data. After the encoding, PQC with unitary gates exploits the spatial information and quanvolve the spatial data like the convolution layers in classical CNN. By measuring the output of PQC, QCNN achieves the quanvoloved data. Note that, by selectively measuring the qubit of PQC, QCNN can reduce the dimensionality like the pooling layers of CNN.

\subsubsection{Encoder} 
The encoder in quanvolutional filter encodes classical information $\mathbf{x}$ into the state information of qubits. There are many encoding strategies, \textit{i.e.}, basis encoding, amplitude encoding, and angle encoding~\cite{alchieri2021introduction}.
In angle encoding, each classical data is encoded via a unitary gate as follows,
\begin{equation}
    \ket{\psi_x} = \prod_{m=0}^{\text{size}(\mathbf{x})} U(\theta_m)U(\mathbf{x}_{m})\ket{0}^{n_q},
\end{equation}
where $U(\theta_m)$ denotes each unitary encoding gate of classical data $m$-th set of data $\mathbf{x}_m$. Angle encoding is generally used in many QCNN-based models due to the simplicity that the angle encoding can encode classical data only with a single rotation. In addition, inspired by ~\cite{perez2020data}, angle encoding with data re-uploading can encode multiple classical data iteratively as follows, 
\begin{equation}
    |\psi_\mathbf{x}\rangle = \prod_{m=0}^{ \ceil{ \text{size}(\mathbf{x}) / n_q} }  U(\theta_m)U(\mathbf{x}_{n_q m:n_q(m+1)})|0\rangle^{n_q}, 
\end{equation} where $\text{size}(\mathbf{x})$ and $\mathbf{x}_{n_q m:n_q(m+1)}$ denote the vector size of input vector $\mathbf{x}$ and the vector which is composed by the first $n_q \cdot m $ to $n_q \cdot (m+1)$ elements, respectively.

\subsubsection{Parameterized Quantum Circuit} As with the universal approximation theorem, there is always a PQC that can represent the optimal objective function within a small error~\cite{benedetti2019parameterized}. Accordingly, QCNN uses PQC with trainable parameters as a filter and proposes to train the filter according to each data-driven task. Using the \textit{chain rule} and \textit{parameter shift rule}, we can write the derivative of the loss function in terms of the output of PQC~\cite{wierichs2022general}.

\subsubsection{Measurement}
The measurement procedure of QCNN is the same as the QML. The output state of each n-qubit filter after training on PQC, $|\psi_l\rangle$, can be measured by a set of projection matrices $\mathbf{M}_i$, which form an orthogonal set,~\textit{i.e.}, $\mathbf{M}_i\mathbf{M}_j=0$ where $i\neq j$. If the set satisfies the requirement $\sum_i \mathbf{M}_i =\mathbf{1}$, 
then any observable $\mathbf{M}$ has the spectral decomposition $\mathbf{M}=\sum_i i\mathbf{M}_i$. The possible outcomes correspond to the eigenvalues $i$ of $\mathbf{M}$. In this paper, we use the measurement operator $\mathbf{M}=\mathbf{I}^{\otimes i-1}\otimes \mathbf{Z} \otimes \mathbf{I}^{\otimes n-i}$, where $\mathbf{I}$ denotes the $2\times 2$ identity matrix and $\mathbf{Z} = \begin{bmatrix}
1 &  0 \\
0 &  -1
\end{bmatrix}$. Then, the measurement can be denoted as, 
\begin{equation}
\langle O \rangle =\langle \psi_{l} | \mathbf{M} |\psi_l\rangle.
\end{equation}

By pooling and fully connecting the measurement $\langle O \rangle$ on the FCN, the QCNN can classify the image classes.

\section{Scalable Quantum CNN}\label{sec:3}
\subsection{Architecture of sQCNN}
Fig.~\ref{fig:Model ARchi} illustrates the architectural difference between QCNN and sQCNN. In classical QCNN with Vanilla-Train, each pixel data is encoded in each qubit. After feature extraction, we can obtain a feature corresponding to the measurement value of each qubit of PQC. Note that, the number of features extracted in this process is equal to the number of qubits in PQC, which is the same as the size of the kernel. Suppose a single quanvolutional layer and a square filter with length $M$. In this case, the number of features (\textit{i.e.,} equal to the number of channels) extracted through the QCNN is fixed to $n_q = M^2$, where $n_q$ is the number of qubits in the filter. In contrast to QCNN, we design sQCNN to be able to increase the number of filters. As the increasing number of qubits in single quantum circuit results in barren plateaus, sQCNN aims to prevent the barren plateaus by increasing the number of filters (\textit{i.e.}, the number of quantum circuits) instead of the number of qubits in the circuit. The sQCNN utilizes multi-filters, enabling sQCNN to adjust the number of extracted features. The number of features is denoted as $n_{f}n_{q}$, where $n_{f}$ is the number of filters. 

\begin{algorithm2e}[t]\label{alg: RF-train}
%\small
    \SetCustomAlgoRuledWidth{0.44\textwidth}  
\caption{Reverse Fidelity Train (RF-Train)}\label{alg:Training algorithm} 
\textbf{Initialization.} \texttt{sQCNN} parameters, $w$;

 \For{ $e = \{1,2,\dots, E\}$}
 {
     \For{ $(x, y) \in \zeta^k$}
     {
         \For{$l, l^{'} \in \{1,2,\dots,L-1\}$}
         {  
            Get features with $l$-th and $l^{'}$-th filter\;
            Calculate $\mathcal{L_{RF}}$\;
            Calculate loss gradients\;
            
         }
         Calculate $\mathcal{L}_e^k \leftarrow \mathcal{L}_{total}$\;
         $\thetab^k_{e+1}\leftarrow \thetab^k_{e}-\eta_e\nabla_{\theta^k_{e}}\lcal^{k}_{e}$\;
     }    
 }
\end{algorithm2e}

\begin{table}[t]
\scriptsize
\centering
\caption{Top-1 accuracy and Euclidean distance comparison }
\resizebox{\columnwidth}{!}{
\begin{tabular}{c||ccc}
    \toprule[1pt]   
     &\multicolumn{3}{c}{\textbf{RF-Train}}\\
    \textbf{Metric} & $\lambda=0$ & $\lambda=0.1$   &  $\lambda=0.5$ \\\midrule
    Top-1 accuracy (\%) & $76$& $78$& $\mathbf{82}$ \\
    Euclidean distance ($\times 10^{-2}$) & $0.4$& $0.7$& $\mathbf{1.1}$ \\
\bottomrule[1pt]
\end{tabular}}\label{tab:distance}
\end{table}

\begin{table}[t!]
\small
    \caption{List of simulation parameters.}
    \centering
    \begin{tabular}{l|r}
    \toprule[1pt]
      \bf{Description}                & \bf{Value}  \\ \midrule
        \# of filters       & \{1, \textbf{2}\} \\
        Optimizer                  & Adam \\
        Initial learning rate      & $10^{-4}$ \\
        \# of qubits in QCNN \& sQCNN               & 4\\
        \# of params in a {QCNN \& sQCNN filter}    & $48$\\
        \# of params in a {classical CNN filter}    & $50$\\
        Kernel size    & {$2^2$}\\
        \bottomrule[1pt]
    \end{tabular}
    \label{tab:tab_parameters}
\end{table}

 \begin{figure*}[t!]
    \centering
    \begin{tabular}{@{}c@{}c@{}}
         \includegraphics[width=.70\columnwidth]{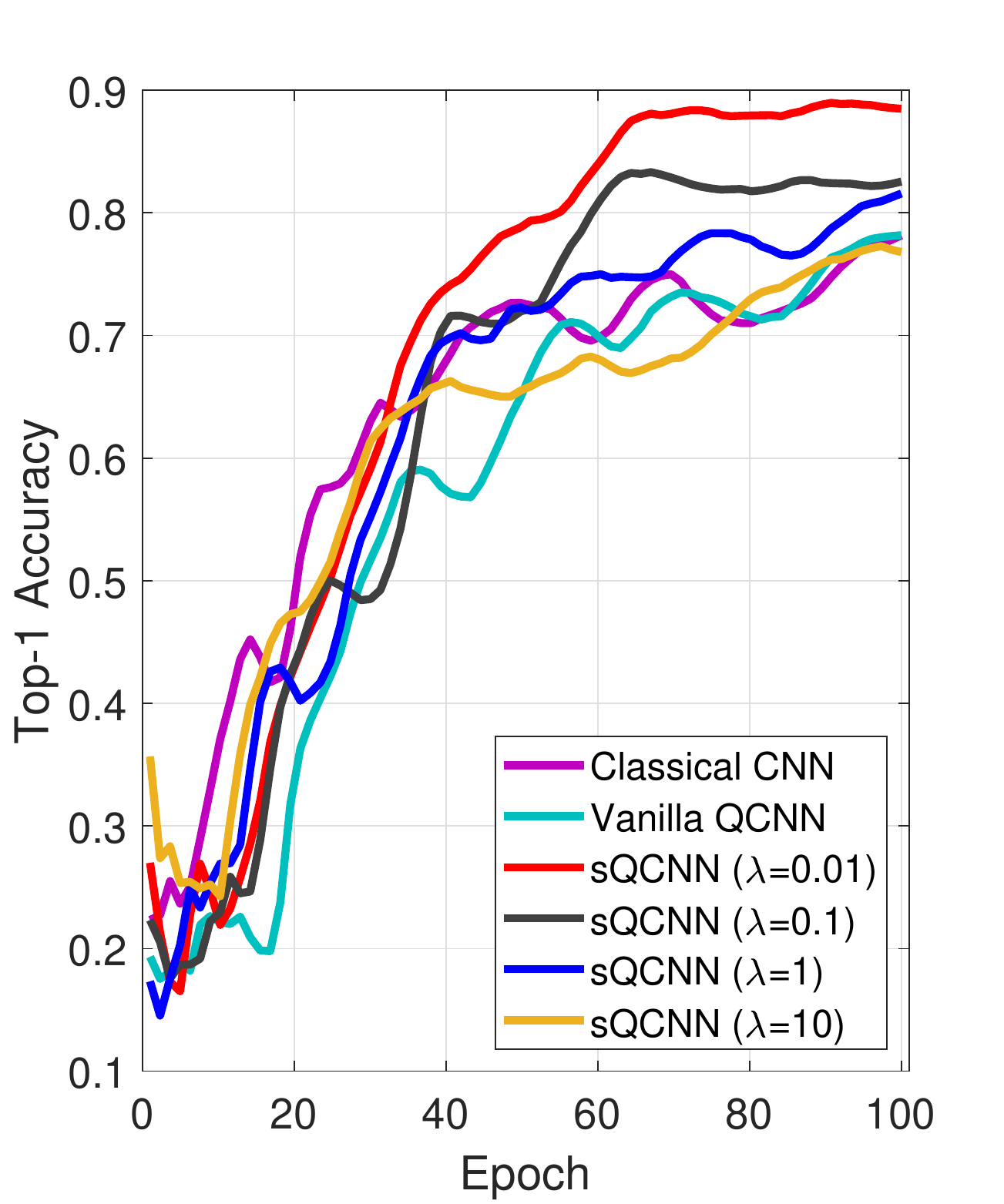}&
         \includegraphics[width=.70\columnwidth]{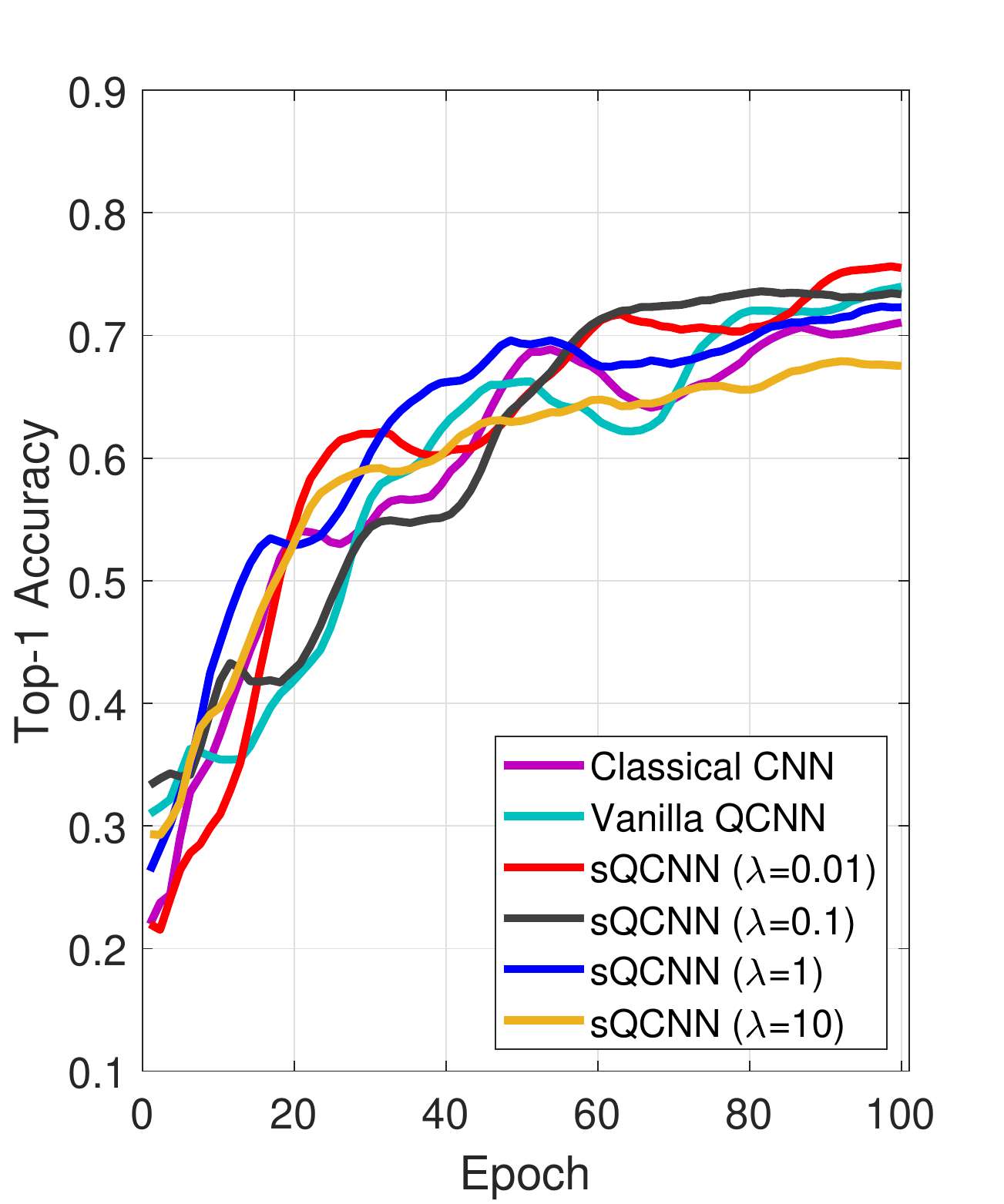}\\
         \small (a) MNIST (10). & \small (b) FMNIST (10).\\
    \end{tabular}
    % \vspace{-3.5mm}
    \caption{Top-1 accuracy of sQCNN with two datasets.}
    \label{exp:2dmain}
\end{figure*}

\begin{figure*}[t]
    \centering
    \includegraphics[width=1.4\columnwidth]{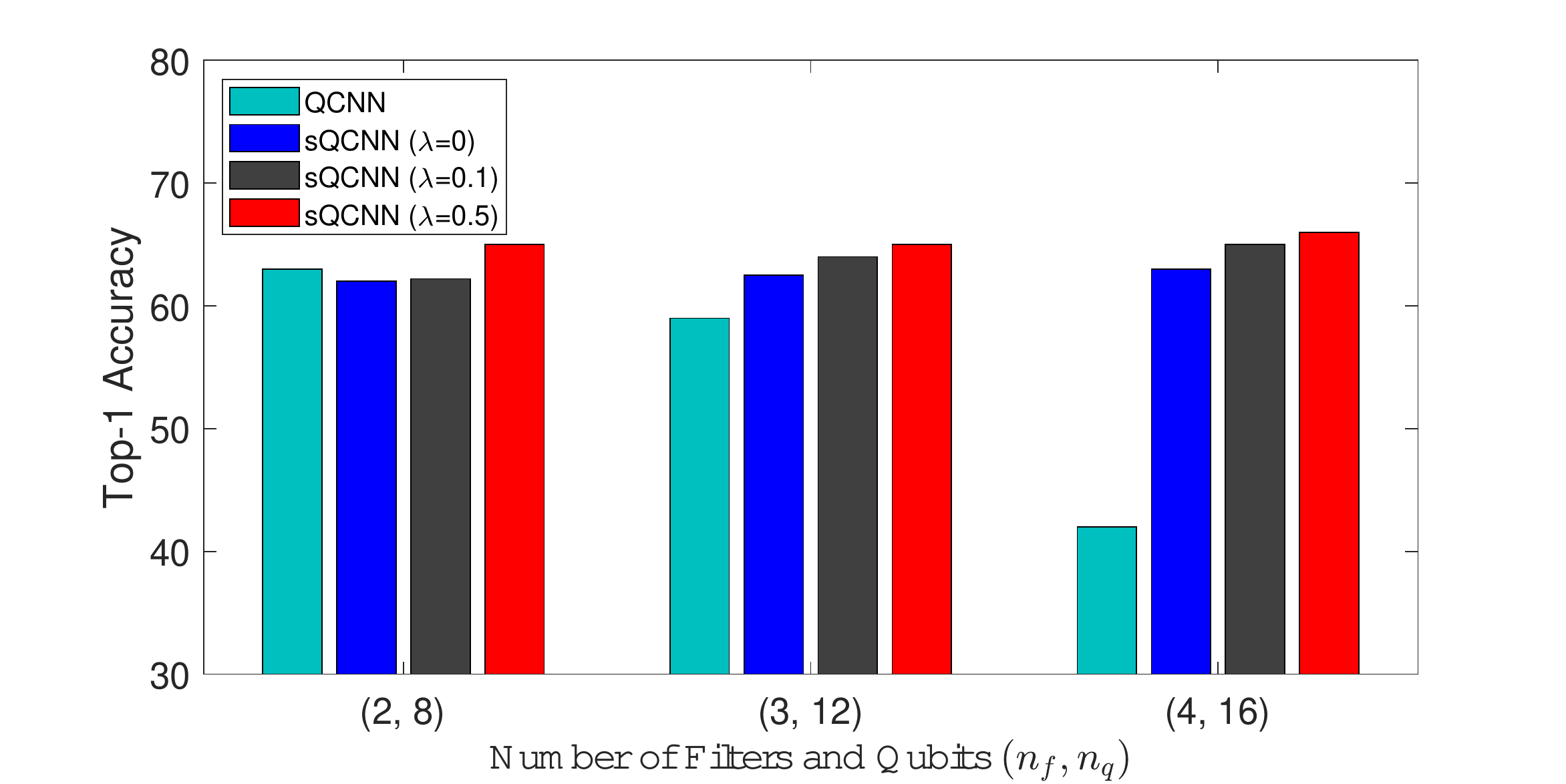}
    \caption{Top-1 accuracy on FMNIST dataset according to the number of filters (qubits). sQCNN ($\lambda = 0, 0.1$ and $0.5$) with filters ($n_f = 2, 3$ and $4$) and QCNN with number of qubits ($n_q =8, 12$ and $16$). }
    \label{exp:main3}
\end{figure*}

\subsection{Reverse Fidelity Train (RF-Train)}

As sQCNN can adjust the number of features and measure each PQC-based quanvolutional filter as a set of features, we aim to fully utilize the architectural advantage of sQCNN. Motivated by the fidelity in quantum computing theory, we propose an sQCNN training algorithm named reverse fidelity-train (RF-Train). The RF-Train contains an RF regularizer that adjusts the fidelity between the quanvolutional filters of sQCNN. Note that fidelity is a nature of quantum computing that is a similarity metric between two quantum states. Here, it is possible to measure fidelity because we scale up the sQCNN by increasing the number of filters, not the number of qubits in a single PQC. Suppose two filters have the output states $|\psi_{l} \rangle$ and $|\psi_{l'}\rangle$, respectively. The fidelity between the two quanvolutional filters is denoted as $\Phi(\rho_{l}, \rho_{l'}) = |\langle \psi_{l}|\psi_{l'}\rangle|^2$, where
$\rho_{l} = |\psi_{l} \rangle \langle \psi_{l}| \textrm{ and }
\rho_{l'} =|\psi_{l'} \rangle \langle \psi_{l'}|$. The increasing similarity between the two quanvoltutional filters drives the fidelity to converge to 1. On the other hand, the fidelity converges to 0 when the similarity between the two filters decreases, indicating that the $l$-th filter does not follow the $l'$-th filter.
We assume that a reduction in the fidelity between output states of the quanvolutional filters enables the extraction of various intrinsic features. We corroborate this assumption with numerical experiments in Sec.~\ref{sec:experiment}. We define the RF regularizer as, 
\begin{equation}
    \mathcal{L_{RF}} = 1 - {1\over L(L-1)} \sum_{l=1}^L  \sum^L_{l'\neq l}  \Phi(\psi_{q_{l}}, \psi_{q_{l^{'}}}),
\end{equation}
where $L$ is the number of filters. With the RF regularizer, the training procedure of sQCNN is described in Algorithm 1. The parameters ($\mathbf{x}$, $y$) are denoted as the input data and label, respectively. We adopt cross-entropy as,
\begin{equation}
\mathcal{L_{CE}} = -{1\over C}\sum^C_{c=1}  \log{p(y_{pred} = y_c|\mathbf{x})}, 
\end{equation}
where $C$ represents the number of classes. $y_{pred}$ and $y_c$ are the predicted and actual classes, respectively. Consequently, we design the total loss of sQCNN as,

\begin{equation} 
\mathcal{L}_{total} ={\frac{1}{|\zeta|}} \sum_{(\mathbf{x}, y) \in \zeta}  [\mathcal{L_{CE}} +\lambda \mathcal{L_{RF}}],
\end{equation}\label{eq:eqlabel}
where $\zeta$, and  $\lambda$ denote the minibatch and an RF regularizer parameter, respectively. By fully utilizing the classification loss $\mathcal{L_{CE}}$ and the regularized RF loss $\mathcal{L_{RF}}$, sQCNN can achieve the diverse features from each measurement, and this results in performance improvement. We corroborate this in Sec.~\ref{sec:experiment}.

\section{Performance Evaluation}\label{sec:experiment}
\subsection{Experimental Setting}

To corroborate the performance of the sQCNN with RF-Train, we design the experiments as follows:
\begin{itemize}
    \item We investigate the top-1 accuracy of sQCNN with various RF-regularizer parameters $\lambda$, and QCNN with Vanilla-Train on both MNIST and FMNIST datasets.
    \item To corroborate the impact of the RF regularizer parameter $\lambda$, we investigate the Euclidean distance between the extracted features due to the RF-regularizer parameter $\lambda$.
    
    \item The scalability of sQCNN is proven. In QCNN, an increase in qubits caused a barren plateau, which degraded the overall model performance. In contrast to QCNN, in sQCNN, more filters result in a performance improvement, despite using the same number of qubits in QCNN. Note that the sQCNN shows performance improvement with the increased number of filters. 
\end{itemize}

\subsection{Experimental Results}
\subsubsection{Performance of sQCNN} 
Fig.~\ref{exp:2dmain} (a)/(b) represent top-1 accuracy of various models with two filters on the MNIST and FMNIST datasets, respectively. Analyzing from the architectural point, both Fig.~\ref{exp:2dmain} (a)/(b) represent that RF-Trained sQCNNs outperform classical Vanilla-Trained QCNN. From the results, we confirmed that even using sQCNN, \textit{i.e.}, even without RF-Train, it can achieve performance improvement in a classification task. Moreover, when the RF-regularizer parameter $\lambda$ increases, sQCNN with a finite number of filters ($n_f =2$) shows performance improvement. sQCNN with high RF-regularizer parameter ($\lambda= 0.5)$ achieves $16 \%$ higher top-1 accuracy than sQCNN which does not utilize RF-Train ($\lambda=0$). Here, we observe that by diversifying the filters with RF-Train, we can improve the performance of sQCNN.

\subsubsection{Impact of RF-Train}
Table~\ref{tab:distance} represents the impact of RF regularizer parameter $\lambda$ on the classification performance. We observe that RF-Train increases the Euclidean distance between extracted features and experimentally confirmed that this diversity between features improves the classification performance of sQCNN.

\subsubsection{Scalability of sQCNN} 
Fig.~\ref{exp:main3} represents that sQCNN achieves scalability. In QCNN, the top-1 accuracy decreases significantly as the number of qubits in a filter increases. As the number of qubits increases from 8 to 16, the performance of QCNN drops about $30\%$. In contrast to QCNN, sQCNN shows a stable performance when the number of filters increases. From the result, we corroborate that this scalability of sQCNN can be a significant characteristic in several tasks requiring many filters as well as simple MNIST and FMNIST data with one input channel.

\section{Conclusions and Future Work}\label{sec:5}
This paper proposes a scalable QCNN (sQCNN) architecture and a novel training algorithm (RF-Train) that enables sQCNN to diversify the extracted features. To achieve scalability while avoiding the barren plateaus which occur when the number of qubits in the filter increases, we utilize multiple filters with a finite number of qubits. To extract various features with the filters and maximize the performance of sQCNN, motivated by the quantum theory, we design an RF regularizer using the concept of fidelity. With extensive experiments, we corroborate the diversity of features extracted by using RF-Train and the scalability of sQCNN. 

As future research, the scalability and trainability of our proposed sQCNN in various applications can be analyzed.

\bibliographystyle{IEEEtran}
\bibliography{PTCL}
\end{document}